\RequirePackage{fix-cm}
\documentclass[onecolumn]{svjour3}
\smartqed  
\usepackage{latexsym}
\usepackage{color}
\usepackage{epsfig}
\usepackage{subfigure}
\usepackage{graphicx}
\usepackage{dcolumn}
\usepackage{stmaryrd}
\usepackage{mathrsfs}
\usepackage{pifont}
\usepackage{amsmath}
\usepackage{amssymb}
\usepackage{bm}
\usepackage{amsfonts}

\newcommand{\la}{\langle}
\newcommand{\ra}{\rangle}

\newcommand{\ga}{\gamma}
\newcommand{\Ga}{\Gamma}

\newcommand{\da}{\dagger}
\newcommand{\De}{\Delta}

\newcommand{\si}{\sigma}

\newcommand{\om}{\omega}

\newcommand{\de}{\delta}

\newcommand{\pa}{\partial}

\def\jpb#1{{ J.\ Phys.\ B} {\bf#1}}

\def\pra#1{{ Phys.\ Rev. A\/} {\bf#1}}
\def\prb#1{{ Phys.\ Rev. B\/} {\bf#1}}
\def\pre#1{{ Phys.\ Rev. E\/} {\bf#1}}
\def\prl#1{{ Phys.\ Rev.\ Lett.} {\bf#1}}
\def\pr#1{{ Phys.\ Rev.} {\bf#1}}

\def\pla#1{{ Phys.\ Lett. A\/} {\bf#1}}
\def\rmp#1{{ Rev. \ Mod. \ Phys.} {\bf#1}}

\journalname{Science Bulletin}
\begin{document}

\title{Overview of Quantum Memory Protection and Adiabaticity Induction by Fast-Signal Control}

\author{Jun Jing         \and         Lian-Ao Wu }

\institute{Jun Jing \at
              Institute of Atomic and Molecular Physics and Provincial Key Laboratory of Applied Atomic and Molecular Spectroscopy, Jilin University, Chuangchun 130012, Jilin, China \\
              \email{junjing@jlu.edu.cn}
           \and
           Lian-Ao Wu \at
              Department of Theoretical Physics and History of Science, The Basque Country University (EHU/UPV), PO Box 644, 48080 Bilbao, Spain \\ Ikerbasque, Basque Foundation for Science, 48011 Bilbao \\
            \email{lianao.wu@ehu.es. Author to whom any correspondence should be addressed.}
}

\date{Received: date / Accepted: date}

\maketitle

\begin{abstract}
A quantum memory or information processing device is subject to disturbance from its surrounding environment or inevitable leakage due to  its contact with other systems. To tackle these problems, several control protocols have been proposed for quantum memory or storage. Among them, the fast-signal control or dynamical decoupling based on external pulse sequences provides a prevailing strategy aimed at suppressing decoherence and preventing the target systems from the leakage or diffusion process. In this paper, we review the applications of this protocol in protecting quantum memory under the non-Markovian dissipative noise and maintaining systems on finite speed adiabatic passages without {\it leakage} therefrom. We analyze leakage and control perturbative and nonperturbative dynamical equations including second-order master equation, quantum-state-diffusion equation, and one-component master equation derived from Feshbach PQ-partitioning technique. It turns out that the quality of fast-modulated signal control is insensitive to configurations of the applied pulse sequences. Specifically, decoherence and leakage will be greatly suppressed as long as the control sequence is able to effectively shift the system beyond the bath cutoff frequency, almost independent of the details of  the control sequences which could be ideal pulses, regular rectangular pulses, random pulses and even noisy pulses.


\end{abstract}

\section{Introduction}\label{intro}

Control of quantum processes \cite{Wiseman} such as quantum storage in open quantum system \cite{Breuer} collects a number of separate concepts, and has a variety of manifestations in different areas of physics \cite{Nielsen,Preskill}. Quantum storage is concerned with a quantum state stored in open systems in presence of leakage \cite{WU09} and diffusion processes induced by either the system-environment interaction or inner-coupling between the target storage subspace and the rest of the system space. It is intimately connected with the phenomena of decoherence and disentanglement in quantum mechanics. As a consequence quantum memory or storage may be regarded as an intersection of quantum information theory \cite{Nielsen} and quantum control theory \cite{Wiseman}. The latter investigates the ability of using an external field to store, transfer, and manipulate information in correlated systems. Since it is important to protect quantum information from degradation, quantum storage protocols are naturally and closely related to the decoherence suppression \cite{PDD,Uhrig08,Uhrig09,West,Kurizki04,Gong10,Kurizki10,Tarn11,Kurizki11,Zhang11,Jing12,XBWang12,Wang12} and the enhancement of adiabaticity of quantum states \cite{Born,Messiah,ZYW}. Despite its long history, quantum decoherence and adiabaticity still challenge our comprehension, and continue to provoke our curiosity, as they are involved with various processes in quantum mechanics. Obviously, we have to be very selective in the topics that we will discuss here, and deal with many important aspects of these topics only briefly.

In quantum information processing, a paramount importance to control of open quantum systems is the quest for combating decoherence. Consider a system embedded in its bath consisted of multiple modes with an arbitrary spectrum distribution (spectral density matrix), for example, an atomic or molecular system in electromagnetic fields, the most probable microscopic process is the resonant exchange of quanta between the system and a particular field mode. Probability of this process is much higher than the Raman scatterings including Stokes and anti-Stokes effects, when the electromagnetic fields are not strong. Thus the leakage rate of the system can be measured by the overlap between the spectra of the pulse-modulated system and that of the bath. In other words, the control quality is determined by the effective gap between the system energy level and the bath cutoff frequency. The system leakage is subject to the accumulation effect during the history that the system exchanges quanta with the environment. For a memoryless or structure-less Markov environment, the open system contacts different modes at different instant, which yields a one-way flow of the quantum information and leaves no chance to memorize the information of a quantum state. On the  contrary, a finite environmental memory time allows the information flow to go back to the system to some extend, which will be considerably helpful to the control effort.

Unified approaches that treat the free dynamical evolution and dynamical decoupling \cite{PDD,DD,CP,Uhrig,UDD,HDD,Gong12} on an equal footing are still under development. It is mostly due to the absence of an exact convolutionless differential equation \cite{Breuer} for the system density matrix, which accommodates arbitrary coupling strength between system and environment. In previous literatures, it seems that in the more convenient way one treats the time evolution of the system, the more unphysical (or unpractical) perturbative method one has to adopt in dealing with the control process. For example, if the control Hamiltonian for a two-level system is described by $H_c=J\si_x$ and the original total Hamiltonian for system and environment is $H_{\rm tot}$, then in order to flip the system in a short time $\de$, the evolution operator combining the control mechanism $U(\de)=-i\si_x\approx\exp(-iH_{\rm tot}\de-iJ\de\si_x)$ requires $J\de=\pi/2$ and $\de\rightarrow0$ \cite{Trotter}. Under this condition, the operation invokes a transverse pulse with infinite strength and zero bandwidth. This amounts to taking the zero-order perturbation for an effective Hamiltonian $H_{\rm eff}=H_{\rm tot}+H_c$ with the negligible perturbation parameter $\de$. Besides the difficulty encountered in experiments, the ideal pulse forces one to ignore the evolution of a strongly-driven system in the finite duration time of the pulse.

The dynamical decoupling and the dynamics of the open quantum system confront with the same difficulty: {\it how to describe the effect from the external environment as well as control on the system dynamics in an exact way?} Microscopically, the system-environment interaction and the environmental statistical property should be carefully taken into the master equation or stochastic Schr\"odinger equation. On the other hand, configurations of the control sequences should be also reflected in the time-dependent system Hamiltonian. One of the first trials along this way was a second-order master equation, with respect to the square of the system-environment coupling strength, targeting on the leakage control of open systems in terms of fast signal control \cite{WU09}. Recently, we proposed a nonperturbative dynamical decoupling \cite{Jing12,Jing13,SR13,SR14} protocol by employing the quantum-state-diffusion (QSD) equation \cite{QSD1,QSD2,QSD3,QSD4,QSD5}, where we can treat the free evolution and control of the system in a united way. Combined with the Feshbach PQ-partitioning technique \cite{WU09}, a one-component differential equation \cite{Jing14} has been derived to address the dynamics of one target instantaneous eigenstate of the system, by which the fast signals control on leakage of quantum memory can be used in manipulating adiabatic condition. Under fast signals control, adiabaticity can be established even when the original Hamiltonian lives in a nonadiabatic regime.

In what follows, we would present a comprehensive revisit to our fast signal control protocol based on the perturbative and nonperturbative dynamical equations. This short review distinguishes itself from any control based on the artificial dynamical decoupling method using the ideal pulse sequence consisted of multiple delta functions in time course. It is clear that this approach is able to eliminate the decoherence and leakage more efficiently with less or optimized control pulses.

\section{Control Equations and Numerical Simulation}

Dynamical decoupling is also termed as bang-bang control in its early days \cite{DD}. As an open-loop method, it is a close cousin of the spin-echo effect \cite{SE}. The decoherence-countering strategies rely on the ability to apply strong and fast pulses. Suppose one needs to eliminate the decoherence effect induced by an operator $X$ in the system-environment interaction Hamiltonian $H_{I}$, one can insert a system operator $A$ satisfying $\{A, X\}=0$ into the time evolution operator $U\equiv\exp(-iH_It)$. Under the short-time approximation, $e^{-i(A+X)t}\approx e^{-iAt}e^{-iXt}$ according to the Trotter formula \cite{Trotter}. It is straightforward to realize a gate $R=e^{-iAt}$ generated by $A$, which will remove the undesired unitary evolution $U_X=e^{-iXt}$ via $RU_XR=U_X^\da$. Therefore, ideal dynamical decoupling or bang-bang control is realized by using the parity-kick cycle. While proposals to control decoherence by realistic (nonideal) pulses have to invoke techniques resulting in master equation up to the second order in the system-bath coupling or stochastic Schr\"odinger equation, which allows to present the shapes of nonideal pulse sequence. That constitutes the main content of this section.

\subsection{Second-order Master Equation}

The most general total Hamiltonian in the framework of open quantum system is represented by the summation of system, bath and their interaction
\begin{equation}
H_{\rm tot}=H_S+H_b+H_I=H_0+H_I,
\end{equation}
where the system-bath interaction term can be always decomposed into $H_I=\sum_jA_jB_j$. $A_j$'s ($B_j$'s) are operators in the space of system (environment). Based on the Nakajima-Zwanzig¡¯s projection \cite{Kurizki04,Alicki02}, $\mathcal{P}[\rho(t)]={\rm Tr}_b[\rho(t)]\otimes\rho_b(t)$, the master equation for the entire system density matrix in the interaction representation (setting $\hbar=1$ in the whole review) is
\begin{equation}\label{MS}
\pa_t\rho_S(t)=-2{\rm Re}\left\{\sum_{mn}\int_0^tdt'
C_{mn}(-t')[A_m(t), A_n(t-t')\rho_S(t)]\right\},
\end{equation}
to the second-order with respect to $H_I$, where $C_{mn}(-t)\equiv{\rm Tr}_b[\rho_bB_mB_n(-t)]$ constitutes the matrix of the bath correlation functions for multi-term system-bath interactions. Here $X(-t)\equiv e^{-iH_0t}Xe^{iH_0t}$, $X=A_j$ or $B_j$. The associated super-operator $\mathcal{P}$ can be redefined as $\mathcal{P}[\cdot]=P\rho_S(t)P\otimes\rho_b$, where $P$ denotes a projection operator onto a desired subspace of the entire system space, in order to study the dynamics and control of the $P$ subspace, then the well-known master equation (\ref{MS}) becomes
\begin{equation}\label{MS2}
\pa_tP\rho_S(t)P=-2{\rm Re}\left\{\sum_{mn}\int_0^tdt'
C_{mn}(-t')P[A_m(t), A_n(t-t')P\rho_S(t)P]P\right\}.
\end{equation}
Suppose the system starts from a pure state $|\phi\ra$, which is one of the orthonormal basis elements constituting the subspace of  $P\equiv|\phi\ra\la\phi|$. Then the probability that the system occupies the $P$-subspace $b(t)$ satisfies
\begin{equation}\label{bt}
\dot{b}(t)=-2b(t){\rm Re}\left[\sum_{mn}\int_0^tdt'
C_{mn}(-t')\mathcal{A}_{mn}(t,t-t')\right],
\end{equation}
where $\mathcal{A}_{mn}(t,s)\equiv\la A_m(t)A_n(s)\ra_\phi-\la A_m(t)\ra_\phi\la A_n(s)\ra_\phi$.

Note that $b(0)=1$. Thus, as a functional of $H_S$, $b(t)$ is equivalent to the fidelity of the system during the leakage process. Inside the integral of Eq.~(\ref{bt}), the only term that could be under control is $\mathcal{A}_{mn}(t,s)$ or $A_m(t)$. Within the framework of the second-order master equation, one can seek a function $H_c(t)$ in the system space, to minimize $|\dot{b}(t)|$ under realistic constraints on pulse energy and width. Physically, Stark shifted by the alternating field might be a straightforward option, e.g. $H_S\Rightarrow H_S+c(t)H_S$. Consider the extreme case that $c(t)=c>0$, when the enhanced system frequency is larger than the cutoff frequency of the bath, $\mathcal{A}_{mn}(t,t-s)$ oscillates faster than the alternating rate of the bath correlation function $C_{mn}(-s)$. The integral of the product of these two functions would oscillates around zero so that $|\dot{b}(t)|$ vanishes. The goal of the leakage control then can be achieved.

Equation~(\ref{bt}) or more explicitly, the fidelity obtained by the second-order master equation
\begin{equation}\label{2dFt}
\mathcal{F}(t)=\exp\left\{-2{\rm Re}\left[\sum_{mn}\int_0^tdt'
C_{mn}(-t')\mathcal{A}_{mn}(t,t-t')\right]\right\}
\end{equation}
is a significant improvement in quantum control methods, in comparison with the ``standard'' bang-bang control, towards an open-loop control method based on the quantum microscopic model. It is a consistent approach, by which the environmental statistical property, i.e. the correlation function and an accessible control pulse sequence, have been taken account into consideration. However, even with the Born approximation (the weak-coupling condition), it is still faraway from satisfaction because it is merely accurate to the second order of system-bath interaction.

\subsection{Quantum-state-diffusion Equation}

Beyond Born and Markov approximations, quantum-state-diffusion equation is capable of dealing with the strong coupling strength and the arbitrary correlation function of the environment. It is also a useful tool in deriving the exact master equation. In addition, the QSD equation obtained with the approximation on the integral over environmental noises may includes the high-order contributions beyond the second-order non-Markovian master equation. It has been found that the QSD equation naturally serves as a useful tool in studying control theories.

Consider a total Hamiltonian describing a quantum system coupled to a bath of bosonic modes:
\begin{equation}
H_{\rm tot}=H_S+\sum_k(g_k^*La_k^\da+g_kL^\da a_k)+\sum_k\om_ka_k^\da a_k,
\end{equation}
where $L$ and $a_k$ ($a^\da_k$) are the coupling operator and annihilation (creation) operator for the $k$-th mode of the bath, respectively. The stochastic wave-function of the system is governed by QSD equation:
\begin{equation}
\pa_t\psi_t(z^*)=-iH_{\rm eff}\psi_t(z^*)=[-iH_S+Lz_t^*-L^\da\bar{O}(t,z^*)]\psi_t(z^*),
\end{equation}
where $z_t^*\equiv-i\sum_kg_k^*z_k^*e^{i\om_kt}$ and $z_k^*$'s are individual Gaussian-distributed complex random numbers. The ensemble average of $z_t^*$ is $M[z_tz_s^*]=\sum_k|g_k|^2e^{-i\om_k(t-s)}$, which is equivalent to $\alpha(t,s)$, the environmental correlation function, at low temperature limit. Note that the ans\"atz $\bar{O}(t,z^*)\psi_t\equiv\int_0^tdsM[z_tz_s^*]O(t,s,z^*)\psi_t$ is a polynomial function of operators acting on the system Hilbert space, which is determined by $\pa_tO(t,s,z^*)=[-iH_{\rm eff}, O(t,s,z^*)]-L^\da\frac{\de\bar{O}(t,z^*)}{\de z_s^*}$. For a given model, a time-local exact QSD equation can be obtained once the exact O-operator is attained. More importantly for control, the formal ans\"atz of QSD equation is irrespective of the Stark shift in the system Hamiltonian $H_S(t)$, which means that the free or control dynamics can be treated on an equal footing. By using Novikov theorem, the QSD equation becomes the following exact form
\begin{equation}
\pa_t\rho_S(t)=-i[H_S, \rho_S(t)]+[L, M[|\psi_t(z^*)\ra\la\psi_t(z^*)|\bar{O}^\da(t,z^*)]]+h.c.
\end{equation}
It will be cast into a convolutionless form in case of $O(t,s,z^*)=O(t,s)$:
\begin{equation}
\pa_t\rho_S(t)=-i[H_S, \rho_S(t)]+[L, \rho_S\bar{O}^\da]+[\bar{O}\rho_S, L^\da].
\end{equation}

To determine the survival probability of the initial state $|\phi\ra$, the fidelity is
\begin{equation}
\mathcal{F}(t)=\la\phi|\rho_S(t)|\phi\ra
=M[\la\phi|\psi_t(z^*)\ra\la\psi_t(z^*)|\phi\ra].
\end{equation}
Different from the general fidelity (\ref{2dFt}), the fidelity based on QSD equation has to be derived case by case. Yet a compensable advantage of QSD equation is that the ``bath'' here (usually referred as a thermal bath with infinite number of modes) can be composed of arbitrary number of modes, which makes it a genuine ``environment''. For example, in the dissipative process of a two-level system $H_S=\frac{E(t)}{2}\om_z$, when the initial state is chosen as $|\phi\ra=\mu|1\ra+\nu|0\ra$, $|\mu|^2+|\nu|^2=1$, the fidelity is found to be
\begin{equation}
\mathcal{F}(t)=1-|\mu|^2-(|\mu|^2-2|\mu|^4)e^{-2\int_0^tds{\rm Re}[F(s)]}+2(|\mu|^2-|\mu|^4){\rm Re}[e^{-\int_0^tdsF(s)}],
\end{equation}
where $F(t)\equiv\int_0^tdsM[z_tz_s^*]f(t,s)$ satisfies $F(0)=0$ and $\pa_tf(t,s)=[iE(t)+F(t)]f(t,s)$. In the controlled dynamics, $E(t)=\om+c(t)$, where $\om$ is the bare frequency of the system and $c(t)$ is the control function.

\subsection{Numerical Results}

Based on the second-order master equation and QSD equation, this subsection presents the fidelity dynamics of an open two-level (qubit) system under the fast signal control. A logic extension to the ideal pulse sequence is the rectangular pulse, periodical or non-periodical, where the period, duration time and the strength for each pulse are finite and experimentally feasible. Moreover, we also extend the regular pulse sequences to those with random pulse and even noisy pulse \cite{SR13,SR14}. In what follows, numerical simulations are performed and the results from random, chaotic \cite{chaos} or noisy pulse \cite{Hanggi1,Hanggi2} are obtained by ensemble average. In order to distinguish effects of the different fast signal sequences, we choose the same environmental spectral density function: $\alpha(t,s)=\frac{\Ga\ga}{2}e^{-\ga|t-s|}$, which is termed as Ornstein-Uhlenbeck noise. The Ornstein-Uhlenbeck process is a useful approach to modeling noisy relaxation with a finite environmental memory time scale $1/\ga$. When $\ga\rightarrow\infty$, the environment memory time approaches to zero and $\alpha(t,s)$ reduces to $\Ga\de(t-s)$, corresponding to the Markov limit.

\begin{figure}[htbp]
\centering
  \includegraphics[width=0.75\textwidth]{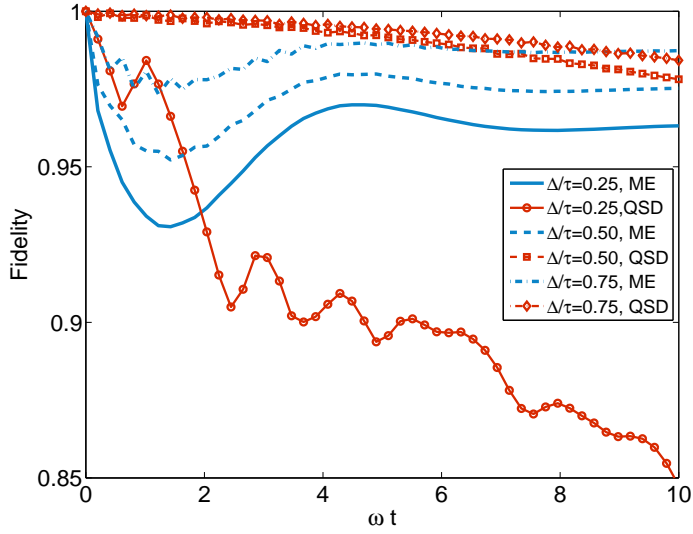}
\caption{(Color online) Fidelity dynamics of a qubit system under regular periodic pulse evaluated with the second-order master equation (ME) and QSD equation. Here we employ the rectangular pulse with period $\tau$, duration time $\De$ and strength $\Psi/\De$, i.e. $c(t)=\Psi/\De$ for regions $n\tau-\De<t\leq n\tau$, $n\geq1$ integer, otherwise $c(t)=0$. The parameters are $\Psi=0.2\omega$, $\tau=0.02\om t$, $\Ga=\omega$, and $\gamma=0.5\omega$. The initial state of qubit satisfies $|\mu|^2=0.5$.}
\label{MEQSD}
\end{figure}

\begin{figure}[htbp]
\centering
  \includegraphics[width=0.75\textwidth]{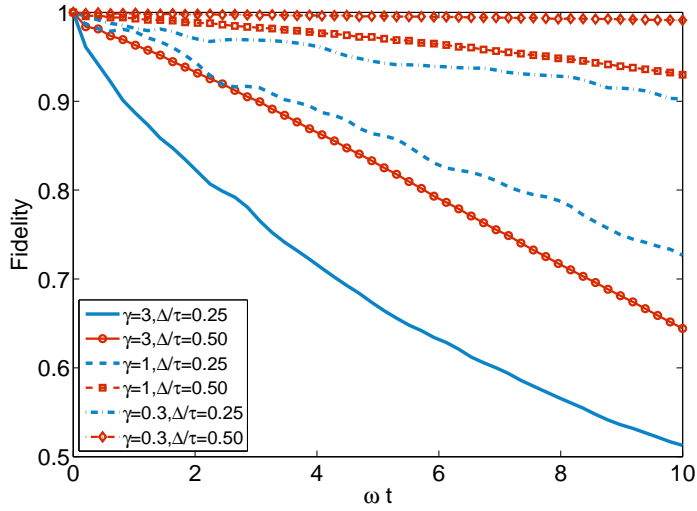}
\caption{(Color online) Fidelity dynamics of a qubit system under random pulse sequence evaluated with QSD equation. Here the random pulse sequence modifies the regular pulse by $X'=X+D_X{\rm Rand}(-1,1)$, where $X=\tau,\De,\Psi$, respectively, $D_X$'s are their individual deviation scales and ${\rm Rand}(-1,1)$ denotes a random number uniformly distributed between $-1$ and $1$. The parameters are $\Psi=0.2\omega$, $\tau=0.02\om t$, $D_\tau=D_\De=0.2\tau$, $D_\Psi=0.9\Psi$, and $\Ga=\omega$. The result has been averaged over ensemble as well as different initial states.}
\label{gamma}
\end{figure}

\begin{figure}[htbp]
\centering
  \includegraphics[width=0.75\textwidth]{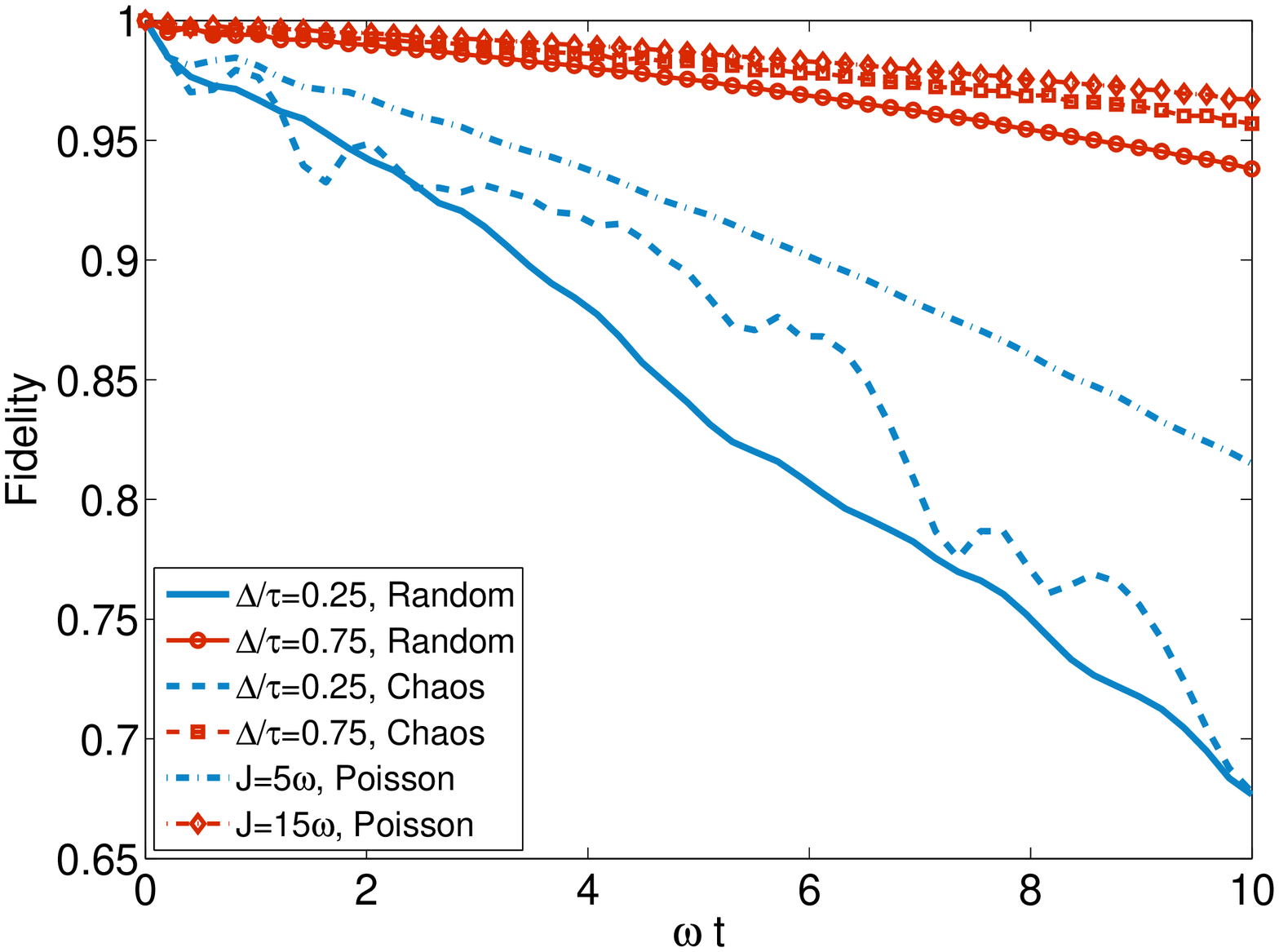}
\caption{(Color online) Fidelity dynamics of a qubit system under different shapes of pulse sequence, including random pulses, chaotic pulses and Poisson white shot noise, evaluated with QSD equation. The random pulse and chaotic pulse are obtained by modifying the regular pulse with strengthes randomly distributed within $[0, \Psi]$ and multiplied with chaotic dimensionless intensity $L_n$, which constitutes a logistic map $L_{n+1}=\mu(L_n-L_n^2)$ with $\mu=3.9$, respectively. The Poisson noise $c(J,W,t)$ satisfies $M[c(J,W,t)]=JW$, where $J$ is the noise strength and $W$ measures the average frequency of noise shots. The parameters are $\Psi=0.4\omega$, $\tau=0.02\om t$, and $\Ga=\omega$. The result has been averaged over ensemble as well as different initial states.}
\label{Noise}
\end{figure}

Figure \ref{MEQSD} demonstrates differences arising from the above two dynamical equations, which decreases with the intense of fast signals. It is shown that under a moderate non-Markovian environment, $\ga=0.5\om$, the second-order master equation cannot precisely describe the control dynamics of the open system. Therefore, the higher-order contribution cannot be omitted in such an environment. Independent of the chosen control parameters, there is always an abnormal interval in the time domain, where the fidelity yields a revival pattern. With both approaches, the control quality under rectangular periodic sequence is steadily enhanced by increasing the ratio of duration time and period of pulse. According to the exact result given by QSD equation, when $\De/\tau\leq0.50$, the fast pulse can maintain the state up to $\mathcal{F}(\om t=10)>0.95$ (see the red dashed and dot-dashed lines).

In the fast signal control protocol with practical pulse sequence, regular pulses can be replaced by random pules, meaning that all the three parameters of the pulses are allowed to stochastically fluctuate around their average values. Physically, it represents the influence from the out-of-control factors or noise resources in laboratory. In Fig. \ref{gamma}, the fluctuation amplitude for both period and duration time is assumed to be $20\%$ and that for the strength is relaxed up to  $90\%$. It is found that under such a remarkable fluctuation, the random control still works as well as regular control, especially under a strong non-Markovian environment. In a near-Markovian environment, $\ga=3$, the time course that the fidelity is preserved above $0.90$ is less than $\om t=3$ even with an intensive fast signal, $\De/\tau=0.50$ (see the red solid line). Yet when $\ga=0.3$, a weak signal control, $\De/\tau=0.25$, can even maintain the fidelity above $0.90$ until $\om t=10$ (see the blue dot-dashed line).

Figures \ref{MEQSD} and \ref{gamma} suggest that the key element of nonideal pulse influencing the fidelity of the open system is the pulse intensity. This result can also be justified by Fig. \ref{Noise}, where we compare the random control (solid lines), the chaotic control (dashed lines) and even the noisy control (dot-dashed lines). Specifically, the strengthes per period along the pulse sequence are no longer identical but with random, chaotic and noisy distributions, respectively. It is shown that as long as the fast signal is sufficiently intensive, $\De/\tau\geq0.75$, the fidelity remains $0.95$ for all the three control sequences at least during the time course $\om t\leq10$. There is no obvious difference among these dynamics. This conclusion not only relaxes the requirement in practical experiments, but also indicates that one should reconsider the underlying reasons for some working and popular dynamical decoupling schemes.

\section{Control and Adiabatic Process}

Although the adiabatic principle had been proposed at the very beginning of the quantum theory, adiabatic passage \cite{AP1,AP2,AP3,AP4} has been reinforced in recent developments in quantum information processing and quantum control. However, in the open quantum system framework, adiabaticity will be often modified and even ruined by the environment noise. In this section, we show that the fast signal control can be used to enhance adiabaticity even induce adiabaticity from a nonadiabatic regime. Contrary to intuition, the transition occurring between different eigenstates can be suppressed not only by an ordered pulse sequence, but also by the chaotic and noisy signals under conditions. To put the protocol into perspective, we present the adiabatic dynamics of one target instantaneous eigenstate by a one-component integro-differential equation based on the following Feshbach PQ partitioning technique.

In general, the wave-function and the effective Hamiltonian in the Schr\"odinger or stochastic Schr\"odinger equation can be always partitioned into,
\begin{equation}\label{PQ}
|\psi(t)\ra=\left[\begin{array}{c} P \\ \hline Q
\end{array}\right], \quad
H_{\rm eff}=\left(\begin{array}{c|c}
      h & R \\ \hline
      W & D
    \end{array}\right),
\end{equation}
according to the interested subspace indicated by $P$ and the irrelevant part $Q$, where the system is prepared at $P$-subspace, i.e. $P(0)=1$ and $Q(0)=0$. In  Eq.~(\ref{PQ}), $h$ and $D$ correspond to the self-Hamiltonians living in the $P$ subspace and the $Q$ subspace, respectively; and $R$ and $W$ are their mutual correlation terms. For the closed system, $R=W^\da$. Consequently, we have
\begin{equation}\label{PE}
i\partial_tP=hP+RQ, \quad i\partial_tQ=WP+DQ.
\end{equation}
The formal expression for $P(t)$ can be rewritten as
\begin{eqnarray}\label{ME}
\pa_tP(t)&=&-ih(t)P(t)-\int_0^tdsR(t)G(t,s)W(s)P(s),
\end{eqnarray}
where $g(t,s)$ incorporates the influence from the remain subspace of the system and external control field and $G(t,s)=\mathcal{T}_{\leftarrow}\{\exp[-i\int_s^tD(s')ds']\}$ is a time-ordered evolution operator. The merit of Eq.~(\ref{ME}) is that it addresses the dynamics of one target component rather than multiple variables. Meanwhile $h(t)$ can be exploited for state control.

Before using Eq.~(\ref{ME}), we now rewrite the Schr\"odinger equation $i\pa_t|\psi(t)\ra=H(t)|\psi(t)\ra$ into the adiabatic representation. The instantaneous eigenequation is $H(t)|E_n(t)\ra=E_n(t)|E_n(t)\ra$, where $E_n(t)$'s and $|E_n\ra$'s are instantaneous eigenvalues and non-degenerate eigenvectors, respectively. A state at time $t$ can then be expressed as $|\psi(t)\ra=\sum_n\psi_n(t)e^{i\theta_n(t)}|E_n(t)\ra$, where $\theta_n(t)\equiv-\int_0^tE_n(s)ds$ is the dynamical phase. Substituting them into the Schr\"odinger equation, we obtain the following differential equation,
\begin{equation}\label{cm}
\pa_t \psi_m=-\la E_m|\dot{E}_m\ra \psi_m-\sum_{n\neq m}\la E_m|\dot{E}_n\ra e^{i(\theta_n-\theta_m)}\psi_n.
\end{equation}
Without loss of generality, the target component can be chosen as $\psi_0$, the amplitude of the target eigenstate $|E_0(t)\ra$ of $H(t)$. Equation~(\ref{cm}) can be regarded as the Schr\"odinger equation for the vector $|\psi(t)\ra=(\psi_0, \psi_1, \psi_2, \cdots)'$ with the effective ``rotating representation" Hamiltonian with $H_{mn}=-i\la E_m|\dot{E}_n\ra e^{i(\theta_n-\theta_m)}$.

Using Eq.~(\ref{ME}), $\psi_0(t)$ satisfies the following one-dimensional integro-differential equation,
\begin{equation}\label{ME1}
\pa_t\psi_0(t)=-\la E_0|\dot{E}_0\ra\psi_0(t)-\int_0^tdsg(t,s)\psi_0(s), \quad g(t,s)=R(t)G(t,s)W(s).
\end{equation}
In this case, $R\equiv[R_1, R_2, \cdots]$ with $R_m=-i\la E_0|\dot{E}_m\ra e^{i(\theta_m-\theta_0)}$, and $W=R^\dag$. The first term on the right-hand side of Eq.~(\ref{ME1}) is the same as that in Eq.~(\ref{cm}), which corresponds to the Berry's phase \cite{Berry1,Berry2} that may be switched off in a rotating frame. $|\psi_0(t)|^2$, the probability of finding the eigenstate $|E_0(t)\ra$ at time $t$, is determined by the accumulation history of product of the propagator $g(t,s)$ and $\psi_0(s)$.

\begin{figure}[htbp]
\centering
  \includegraphics[width=0.75\textwidth]{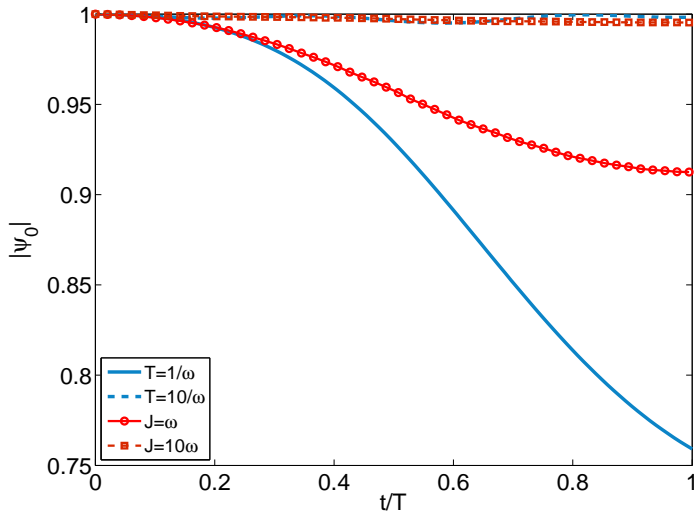}
\caption{(Color online) $|\psi_0|$ vs dimensionless time $t/T$ for different passage times $T$, which is the key parameter in a widely used model in quantum adiabatic algorithms: $H(t)=\om[t/T\si_x+(1-t/T)\si_z]$. The adiabatic limit $|\psi_0|\approx1$ is achieved either for larger $T$ when the system follows an adiabatic path or with the assistance by Poisson white shot noise.}
\label{AdiaC}
\end{figure}

With the exact dynamical equation~(\ref{ME1}), a crucial and general adiabatic condition can be cast into the following compact form,
\begin{equation}\label{ME2}
\int_0^tds\,g(t,s)\psi_0(s) =0.
\end{equation}
The condition is satisfied when $g(t,s)=0$ or $g(t,s)$ is factored into a product of  one rapid oscillating function around zero and one much slowly varying function. Mathematically, it is understood that the integral of the product of the fast-varying $g(t,s)$ and the slow-varying $\psi_0(s)$ gives rise to a vanishing result. For a two-level system with frequency difference $E(t)=E_0-E_1$, when it is initially prepared as the eigenstate $|E_0\ra$, the propagator $g(t,s)$ is given by,
\begin{equation}\label{Pt}
g(t,s)=-\la E_0(t)|\dot{E}_1(t)\ra\la E_1(s)|\dot{E}_0(s)\ra e^{\int_s^t(iE-\la E_1|\dot{E}_1\ra)ds'}.
\end{equation}
If $E(t)$ can be manipulated by fast signal, then the exponential function in $g(t,s)$ will play a crucial role to make the absolute value of the integral in Eqs.~(\ref{ME1}) or (\ref{ME2}) as small as possible.

In Fig. \ref{AdiaC}, we consider an time-dependent Hamiltonian that is of those typical models describing adiabatic passage: $H(t)=a(t/T)H_1+b(t/T)H_2$, where $a(0)=1$, $b(0)=0$ and $a(T)=1$, $b(T)=0$. $T$ is a key element to observe the adiabatic passage time. It is known for these models, when $T$ is sufficient large, the system state can spontaneously adiabatically evolve from one of the instantaneous eigenvector of $a(t)H_1$ to the corresponding eigenvector of $b(t/T)H_2$. Otherwise, the transition between different eigenstates of $H(t)$ occurs. The two blue lines in Fig. \ref{AdiaC} illustrate the difference between the slow adiabatic and the diabatic passages. The former passage is about $10$ times as long as the latter one. According to Eq.~(\ref{ME1}), if $|\psi_0|$ is maintained as unity, then the adiabatic passage can also be realized. Here the system frequency $\om$ is modulated as $\om+c(t)$, where $c(t)=c(W,J,t)$, the Poisson white shot noise. The red lines in Fig. \ref{AdiaC} show that we can greatly accelerate the passages with the help of fast signal when it is sufficiently intensive.

\section{Conclusion}

In this short review, we systematically report our progress on fast signals control methods in quantum storage and adiabatic process. The nonperturbative control method has been developed for decoherence- and leakage-suppression. The results based on the second-order master equation and QSD equation demonstrate that a system dynamics can be {\it stabilized} in terms of arbitrary configurations of the fast signals. In particular situations, the environmental dissipative noise can even be neutralized by the white noise. The fast signal control can also be used in realizing a shortcut to adiabaticity and to the suppression of the leakage from the target adiabatic passages. Our strategy remarkably relaxes the experimental requirements in precisely-engineering control sequences.

\begin{acknowledgements}
We acknowledge grant support from the Basque Government (grant IT472-10), the Spanish MICINN (No. FIS2012-36673-C03-03) and the NSFC No. 11175110.
\end{acknowledgements}

\end{document}